\documentclass[twocolumn,showpacs,preprintnumbers,nofootinbib,amsmath,amssymb,fleqn]{revtex4}
\usepackage{graphicx, psfrag}
\usepackage{epsfig}
\usepackage{hyperref}

\def\bx{{\bf x}}

\newcommand{\tr}{\mbox{Tr$\:$}} 
\newcommand{\non}{\nonumber}
\newcommand{\del}[2]{\mbox{$\delta^{#1#2}$}}   
\def\lsim{\raise0.3ex\hbox{$<$\kern-0.75em\raise-1.1ex\hbox{$\sim$}}}
\def\gsim{\raise0.3ex\hbox{$>$\kern-0.75em\raise-1.1ex\hbox{$\sim$}}}

\begin{document}

\title{
Heavy quark free energies for three quark systems at finite temperature
}

\author{Kay H\"{u}bner}
\author{Frithjof Karsch}
\affiliation{ 
Physics Department, Brookhaven National Laboratory,
Upton, NY 11973, USA
}

\author{Olaf Kaczmarek}
\author{Oliver Vogt}

\affiliation{ 
Fakult\"{a}t f\"{u}r Physik, Universit\"{a}t 
Bielefeld, D-33615 Bielefeld, Germany
}

\date{\today}

\begin{abstract}
We study the free energy of static three quark 
systems in singlet, octet, decuplet and average color channels 
in the quenched approximation and in 2-flavor QCD at finite temperature. 
We show that in the high temperature phase singlet and decuplet free energies of three
quark systems are well described by the sum of the free energies of three
diquark systems plus self energy contributions of the three quarks.
In the confining low temperature phase we find evidence for a Y-shaped
flux tube in SU(3) pure gauge theory, which is less evident in 2-flavor QCD due to the onset of string breaking.
We also compare the short distance behavior of octet and decuplet free energies to the free energies of single static quarks in the corresponding color representations.  
\end{abstract}

\maketitle

\section{Introduction}

The study of baryonic systems composed of three static quarks 
sheds light 
on the internal structure of the baryon. 
Lattice calculations have been carried out addressing this question for quite some time now
\cite{Sommer:1985da,Bali:2000gf,Alexandrou:2001ip,Takahashi:2002bw,Takahashi:2004rw,Bornyakov:2004uv,Bornyakov:2004yg,deForcrand:2005vv}.
Of particular interest is the question
whether a genuine three body force exists between the quarks below the critical
temperature and how the system
behaves at finite temperature. 
At zero temperature the Y-string shape of the color flux tube in baryonic systems is supported by recent
calculations in lattice QCD \cite{Takahashi:2002bw,Bornyakov:2004uv,deForcrand:2005vv}. At finite temperature work
so far has concentrated on simulations using maximal abelian
gauge \cite{Bornyakov:2004uv,Bornyakov:2004yg}, showing the existence of a Y-shaped string as well. 
The Y-Ansatz is also  supported by results obtainde with the field correlator method \cite{Kuzmenko:2000bq}.

Recently, there have been speculations that  hadronic bound states, e.~g.~also colored $QQQ$-states, may still exist in the high temperature deconfined phase of QCD and that they could be responsible for the observed strong interactions in the QGP near $T_c$\cite{Liao:2005hj}. 
This makes it interesting to analyze the force in three quark systems also in non-singlet color configurations that might exist in an overall color neutral thermal medium.

In this paper, we examine the question of the flux tube shape of the
$QQQ$-singlet state below the
critical temperature in the quenched approximation of QCD and discuss the
appearance of string breaking in 2-flavor QCD.
Moreover, above $T_c$ we explore the relation between the free energies in different
color channels of the baryonic system and the free energies of $QQ$-subsystems, and
analyze the screening of octet and decuplet free energies at small distances.

This paper is organised as follows:
In section \ref{Channels} we present the basic observables we use
for calculations in different color channels of the heavy three quark system 
in terms of correlation functions of thermal Wilson lines. 
In section \ref{perttheory} we discuss the perturbative behavior of free energies of three
quark systems at short distances.
In section \ref{details} we present the details of our simulation, including
calculation techniques, gauge fixing and renormalisation procedures. 
In section \ref{puregauge} we discuss the behavior of $QQQ$-free energies obtained in SU(3) pure gauge theory on $32^3\times 4,8$ lattices and compare results above $T_c$
with free energies of $QQ$-systems. Furthermore,  we examine the string shape below $T_c$.
Section \ref{fullqcd} is devoted to a discussion of our results in 2-flavor QCD.
In section \ref{sec:screening} we analyze the screening of octet and decuplet free energies at small distances.  
Finally, in section \ref{conclusion}, we conclude.

\section{Static three Quark Systems in different color representations}
\label{Channels}

In the following we will construct observables for the free energies
of static three quark systems in different color channels and introduce the corresponding relation to the expectation values
of correlation functions of thermal Wilson lines. 
Since we are interested in QCD, we restrict ourselves to examine the SU(3)
gauge group.

The state of a three quark system as the product of irreducable
representations of three quarks in color space can be decomposed into symmetry states 
\begin{equation}
  \label{eq:3q_basis}
  3\otimes 3\otimes 3  =  1\oplus 8\oplus 8^{\prime}\oplus 10.
\end{equation}
%
The singlet is totally anti-symmetric, the first octet anti-symmetric in the
first and second, the second octet in the second and third component and the
decuplet is totally symmetric. 

We now construct observables for static three quark systems represented by suitable combinations of Wilson lines.
The derivation is similiar to
that for two quark systems \cite{Nadkarni:1986as,Nadkarni:1986cz}, but more elaborate.
We start by defining the thermal Wilson line
\begin{equation}
  \label{eq:thermal_wilson}
  L(\bx)=\prod_{x_4=0}^{N_\tau-1}U_4(\bx,x_4).
\end{equation}
The Polyakov loop is then obtained by $l(\bx)=\tr L(\bx)$, where the trace is normalised such that $\tr\mathbf{1} = 3$.
For the three quark correlation function of 
thermal Wilson lines\footnote{For convenience we write $L_i$ instead of $L(\bx_i)$.} $L_1, L_2, L_3$, we have
\begin{equation}
  \label{eq:correlation}
  L_1 L_2 L_3 =\sum_s C_{QQQ}^{s}(\bx_1, \bx_2, \bx_3)P_s,
\end{equation}
%
where $s\in\{1, 8, 8^{\prime}, 10\}$ stands for the symmetry states, $P_s$ for the
projectors on these states and $C_{QQQ}^{s}$ for the three point correlator of the
thermal Wilson line belonging to that symmetry state. Here we have suppressed the
dependence on the temperature $T$.


Denoting the components of the thermal Wilson lines $L_1^{il}, L_2^{jm}, L_3^{kn}$, 
where $i,j,k,l,m,n=1,2,3$, 
we find the projectors $P_s$ by using the Fierz-identity
\begin{eqnarray}
  \label{eq:projectors1}
  P_{1} & = & \frac{1}{6}\sum_{\sigma(l,m,n)}\epsilon^{lmn}
  \del{i}{l}\del{j}{m}\del{k}{n}\\\non\\
  \label{eq:projectors8}
  P_{8}
  &=&\frac{1}{3}\sum_{\sigma(l,m,n)}\epsilon^{lm}
  \del{i}{l}\del{j}{m}\del{k}{n}\\\non\\ 
  \label{eq:projectors8_2}
  P_{8^{\prime}} 
  &=&\frac{1}{3}\sum_{\sigma(l,m,n)} \epsilon^{nm}
  \del{i}{l}\del{j}{m}\del{k}{n}\\\non\\ 
  \label{eq:projectors10}
  P_{10} & = & \frac{1}{6}\sum_{\sigma(l,m,n)}\del{i}{l}\del{j}{m}\del{k}{n},\\\non
\end{eqnarray}
where the sums are over all permutations $\sigma$ of the indices. 
The $P_s$ satisfy the usual relations for projectors 
\begin{equation}
  \label{eq:projector_rel}
  P_{s}^2=P_{s},\quad \sum_s P_s=\mathbf{1}\quad\text{und}\quad P_sP_t=0
\end{equation}
for $s\ne t$ and $s,t\in \{1,8,8^{\prime},10\}$.
The desired three quark correlators in the different color channels are now obtained by applying
\begin{equation}
  \label{eq:projector_rule}
  C_{QQQ}^{s} = \frac{\tr \Big(P_s L_1 L_2 L_3 \Big)}{\tr P_s}.
\end{equation}
%
We find
\begin{eqnarray}
  \non
  \label{eq:sing_frenergie}
  \lefteqn{C_{QQQ}^{1}}\\\non
   &=&\frac{1}{6} \big(27\,\tr L_1 \tr L_2 \tr L_3 - 9\,\tr L_1 \tr(L_2L_3)\big.\\\non
   &&\quad\big.- 9\,\tr L_2 \tr(L_1L_3)- 9\,\tr L_3 \tr(L_1L_2)\big.\\
   &&\quad\big.+ 3\,\tr(L_1L_2L_3) + 3\,\tr(L_1L_3L_2)\big)\\\non\\
   \non
   \label{eq:ok1_frenergie}
   \lefteqn{C_{QQQ}^{8}}\\\non
   & = &\frac{1}{24}\big(27\,\tr L_1 \tr L_2 \tr L_3 + 9\,\tr L_1 \tr(L_2L_3)\\
   &&\quad\big.  - 9\,\tr L_3 \tr(L_1L_2)- 3\,\tr (L_1 L_3L_2) \big)\\\non &&\\
   \non
   \label{eq:ok2_frenergie}
   \lefteqn{C_{QQQ}^{8^{\prime}}}\\\non
   & = &\frac{1}{24}\big( 27\,\tr L_1 \tr L_2 \tr L_3 + 9\,\tr L_3 \tr(L_1L_2)\\
   &&\quad  \big. - 9\,\tr L_1 \tr(L_2L_3)- 3\,\tr (L_1 L_2 L_3) \big)\\\non
   &&\\
   \non
   \label{eq:dek_frenergie}
   \lefteqn{C_{QQQ}^{10}}\\\non 
   & = &\frac{1}{60}\big(27\,\tr L_1 \tr L_2\tr L_3 + 9\,\tr L_1 \tr(L_2L_3)\big.\\\non
     &&\quad \big. + 9\,\tr L_2 \tr(L_1L_3) + 9\,\tr L_3 \tr(L_1L_2) \big.\\
     &&\quad \big.+ 3\,\tr(L_1L_2L_3) + 3\,\tr(L_1L_3L_2)\big).\\\non
   \non
\end{eqnarray}
%
Finally we obtain for the color averaged correlator of the three quark system,
$C^{\text{av}}_{QQQ}$, the relation
\begin{eqnarray}
\non
  C^{\text{av}}_{QQQ}
  &=&\frac{1}{27}C_{QQQ}^{1}
  +\frac{8}{27}C_{QQQ}^{8}
  +\frac{8}{27}C_{QQQ}^{8^{\prime}}
  +\frac{10}{27}C_{QQQ}^{10}\\\non&&\\ 
  &=&\frac{1}{27}\tr L_1 \tr L_2 \tr L_3.
\label{eq:average_QQQ}
\end{eqnarray}
%
%
We note that the two octet correlators $C_{QQQ}^{8}$ and $C_{QQQ}^{8^{\prime}}$ are the same when calculated on the lattice. 
The free energy of the symmetry state $s$ can be obtained from the
correlator in the usual way
\begin{equation}
  \label{eq:cor_freeenergy}
  F_{QQQ}^{s}(T)=-T\ln \left\langle C_{QQQ}^{s}(T)\right\rangle,
\end{equation}
where $\left\langle\cdot\right\rangle$ stands for thermal averages taken after gauge fixing. We have suppressed the position dependence on both sides of \eqref{eq:cor_freeenergy}. 

In the following section we will discuss briefly the behavior of the free energies of the three
quark system for small couplings and short distances.

\section{${\mathbf F^s_{QQQ}}$ at short distances}
\label{perttheory}

In the perturbative expansion of the free energy of $QQQ$-systems the contribution of the three gluon
vertex vanishes for symmetry reasons \cite{Cornwall:1996xr}. Therefore, neglecting
self energy contributions, to order
$g^4$ the free energy $F^s_{QQQ}$ decomposes into the sum of three diquark
free energies $F^t_{QQ}$, 
which can be in an anti-symmetric antitriplet $(t=\overline{3})$ or in a
symmetric sextet $(t=6)$ state. The behavior of these diquark free energies at small distances to lowest order are Coulombic
 \begin{equation}
   \label{eq:fre_sample_qq}
    F^t(R,T)= C_2(t)\frac{\alpha}{R},
 \end{equation}
where $\alpha=\frac{g^2}{4\pi}$ and  $R$ is the separation of the static quarks in the $QQ$-system.
The Casimir factor  $C_2= \tr\left(t_1^at_2^a\right)$  depends on the color channel $t$ the diquark system is in and can be found in table \ref{prefactor_tab}.  
In order to obey the permutation relations given for the $QQQ$-system in
\eqref{eq:projectors1}  - \eqref{eq:projectors10}, the $QQQ$-singlet state must be composed of $QQ$-anti-triplets, the $QQQ$-decuplet of $QQ$-sextets and the $QQQ$-octets are a mixture of $QQ$-anti-triplets and -sextets. Thus, we have for small distances
\begin{eqnarray}
  \label{eq:fre_QQQ_singlet_small}
  F_{QQQ}^1({\bf R},T) &=&\sum_{i<j}-\frac{2}{3}\frac{\alpha}{R_{ij}}  + \tilde{k}_1(T)\\\non
  &&\\
   \label{eq:fre_QQQ_octet1_small}
	\non
   F_{QQQ}^8({\bf R},T) &=&
    -\frac{2}{3}\frac{\alpha}{R_{12}}
    -\frac{1}{6}\frac{\alpha}{R_{13}}+\frac{1}{3}\frac{\alpha}{R_{23}}+\tilde{k}_8(T)\\\non &&\\
  &&\\
  \label{eq:fre_QQQ_octet2_small}
	\non
   F_{QQQ}^{8^{\prime}}({\bf R},T) &=&\frac{1}{3}\frac{\alpha}{R_{12}}
    +\frac{1}{6}\frac{\alpha}{R_{13}}-\frac{2}{3}\frac{\alpha}{R_{23}}-
    +\tilde{k}_{8^\prime}(T)\\\non &&\\
  &&\\
  \label{eq:fre_QQQ_decuplet_small}
  F_{QQQ}^{10}({\bf R},T) &=&\sum_{i<j} \frac{1}{3}\frac{\alpha}{R_{ij}}+ \tilde{k}_{10}(T),
\end{eqnarray}
where $R_{ij}$ denotes the distance between the $i$th and $j$th quark (see
sec.~\ref{calc_technique} for details), ${\bf R}=(R_{12},R_{13},R_{23})$. The self energy contributions $\tilde{k}_i$ with $i=1,8,8^\prime,10$ are temperature dependent and at small separations related to the free energy of a single static quark in the corresponding color state, $F_Q^{(s)}(T)$ (see sec.~\ref{sec:screening} for details). Thus $\tilde{k}_1(T)=0$, $\tilde{k}_8(T)=\tilde{k}_{8^\prime}(T)=F^{(8)}_Q(T)$ and $\tilde{k}_{10}(T)=F^{(10)}_Q(T)$. 
As can be seen easily from \eqref{eq:fre_QQQ_singlet_small} and
\eqref{eq:fre_QQQ_decuplet_small}, the singlet state of the $QQQ$-system is
attractive, because the $QQ$-antitriplet is, and the decuplet state is repulsive, as
the $QQ$-sextet is.  
Moreover, the $QQQ$-singlet free energy is temperature independent at small distances like the $Q\bar Q$-singlet free energy is \cite{Kaczmarek:2002mc}.
For the octet channels the situation is more complicated due to the contribution of both anti-triplet and sextet free energies.
We can, however, compute the lowest order behavior of the $QQQ$ free energies for equilateral geometries for all color channels.
Table \ref{prefactor_tab} summarizes the Casimir factors for the free energies of the different
color channels for all the quark systems $Q\bar Q$, $QQ$ and $QQQ$. 
One recovers
the lowest order behavior of the free energy in some symmetry state $t$ (equilateral geomtries for the $QQQ$-systems only) that depends on the separation of the static quarks by using (\ref{eq:fre_sample_qq}),
%
%
where $R$ is now the separation of the static quarks in the $Q\bar Q$- and $QQ$-systems and the edge length of the equilateral geometry in the $QQQ$ case. See sec.~\ref{calc_technique} for details on the geometric configuration of the three static quarks.  We see, that the average of both $QQQ$-octet free energies (which is accessible to lattice calculations) for equilateral geometries is expected to be weakly attractive.  

\begin{table}
  \begin{center}
    \begin{tabular}{lccccc}
      &&&&&\\
       system & $1$  & $\overline{3}$ & $6$ & $8$ & $10$\\
       \hline
       $Q\overline{Q}$ &$-4/3$&&&$+1/6$&\\
       $QQ$      &&$-2/3$&$+1/3$&&\\
       $QQQ^{\dagger}$     &$-2$&&&$-1/2^{\star}$&$+1$\\
       \hline
       &&&&&\\
     \end{tabular}
     \caption{\label{prefactor_tab}Casimir factor for color symmetry state $t$, $C_2(t)= \tr\left(T_1^aT_2^a\right)$.
       \\ $\dagger$ equilateral geometries, $^{\star}$ average of both octets.}
   \end{center}
\end{table}

At larger distances, the pairwise interactions of the static quarks should receive temperature dependent contributions and the Coulomb terms in (\ref{eq:fre_QQQ_singlet_small}) - (\ref{eq:fre_QQQ_decuplet_small}) have to be replaced by the full diquark free energies $F^t_{QQ}(R,T)$ of the corresponding color channel $t$.
In this case, (\ref{eq:fre_QQQ_singlet_small}) - (\ref{eq:fre_QQQ_decuplet_small}) generalize to  
\begin{eqnarray}
  \label{eq:fre_QQQ_singlet}
  F_{QQQ}^1({\bf R},T) &=&\sum_{i<j} F_{QQ}^{\overline {3}}(R_{ij},T) + k_1(T)\\\non
  &&\\\non
   \label{eq:fre_QQQ_octet1}
   F_{QQQ}^8({\bf R},T) &=&F_{QQ}^{\overline {3}}(R_{12})+\frac{1}{4}F_{QQ}^{\overline {3}}(R_{13})\\&&+F_{QQ}^6(R_{23})+k_8(T)\\\non
  &&\\\non
  \label{eq:fre_QQQ_octet2}
   F_{QQQ}^{8^{\prime}}({\bf R},T) &=&F_{QQ}^6(R_{12})-\frac{1}{4}F_{QQ}^{\overline {3}}(R_{13})\\&&+F_{QQ}^{\overline {3}}(R_{23})+k_{8^\prime}(T)\\\non
  &&\\
  \label{eq:fre_QQQ_decuplet}
  F_{QQQ}^{10}({\bf R},T) &=&\sum_{i<j} F_{QQ}^6(R_{ij},T) + k_{10}(T),
\end{eqnarray}
where the $k_i(T)$ with $i=1,8,8^\prime,10$ account for the self energy contributions not included in the diquark free energies. 
It was shown in
\cite{Doring:2007uh} that the residual $T$-dependence of the $QQ$-anti-triplet free energy
at small distances in the deconfined phase can be removed by substracting the
free energy of a single static quark
$F_Q(T)\equiv\frac{1}{2}\lim_{R\to\infty}F^1_{Q\bar Q}(R,T)$.   
We therefore expect $k_1(T)=-3F_Q(T)$, which guarantees the cancelation of self energy contributions stemming from $F^{\bar 3}_{QQ}$, leaving $F^1_{QQQ}$ temperature independent at small separations.    
Moreover, at large separations, the self energy contributions for the $QQQ$-system should be independent from the particular color channel the system is in, leading to
 $k_1(T)=k_8(T)=k_{8^\prime}(T)=k_{10}(T)=-3F_Q(T)$.
We will show in
sections \ref{puregauge},\ref{fullqcd} and \ref{sec:screening} that this reasoning is indeed correct for temperatures $T>T_c$.
Indeed, we will show that the decomposition of interactions in 3-quark systems in terms of 2-quark interactions as suggested by (\ref{eq:fre_QQQ_singlet})-(\ref{eq:fre_QQQ_decuplet}) holds for $T>T_c$.

We finally note, that the $QQQ$-octet free energy obtained from the lattice is the average of $F^8_{QQQ}$ and $F^{8^\prime}_{QQQ}$, as will be shown in sec.~\ref{details}. Consequently, we expect the lattice $QQQ$-octet free energy for equilateral geometries to obey 
 \begin{eqnarray}
   \label{eq:fre_eight}
   \non
    F^8_{QQQ}(R,T)&=&\frac{3}{2}\left(F_{QQ}^{\bar 3}(R,T)  + F_{QQ}^6(R,T) \right)\\
    && -3F_Q(T),
 \end{eqnarray}
where $R$ is the edge length of the equilateral triangle.


\section{Simulation Details}
\label{details}

We will now present the details of our simulation.
\begin{figure}[t]
\center{
    \psfrag{a}[c]{$Q_1$}
  \psfrag{b}[c]{$Q_2$}
   \psfrag{c}[c]{$\,\,Q_3$}
  \psfrag{d}[c]{$\quad R_{23}$}
   \psfrag{e}[c]{$\quad R_{13}$}
  \psfrag{f}[c]{$\!\!\!\! R_{12}$}
   \psfrag{p}[c]{$F$}
  \psfrag{w}[c]{$\frac{2\pi}{3}$}
  \epsfig{file=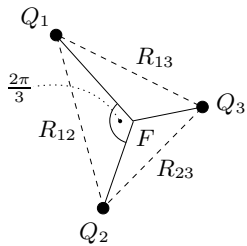, height=3cm}
  \caption{\label{fig_3eck}Inter quark distances. $F$ is the Fermat point of
    the triangle.} 
  
}
\end{figure} 

\subsection{Simulation}
We used gauge field configurations generated on $32^3\times 4$ and $32^3\times 8$ lattices in pure gauge theory
with the tree level-Symanzik improved gauge action at several couplings above and below the
critical coupling \cite{weisz1,weisz2}.   
In 2-flavor staggered QCD we reexamined configurations on 
a $16^3\times 4$ lattice for several different couplings at bare quark mass $m/T=0.4$, where for fermions
the p4-action and for the gauge fields again the tree level-Symanzik improved
gauge action were
used \cite{Karsch:2000kv,Allton:2002zi,Allton:2003vx}. 
The scale was set using the string tension, $\sigma$, at $T=0$ following \cite{beinlich} and use it to express all dimensionful observables.
On all configurations we calculated the three point correlation
functions of Polyakov loops \eqref{eq:sing_frenergie} - \eqref{eq:dek_frenergie},
\eqref{eq:average_QQQ} in a manner explained below. 
%

The operators in \eqref{eq:sing_frenergie} - \eqref{eq:dek_frenergie} are
not manifestly gauge independent and therefore a gauge fixing procedure must be applied to
our gauge configurations.
We use Coulomb gauge for our calculations, in which the singlet free energies of the
$Q\bar Q$-system are related to a gauge independent definition in
terms of dressed Polyakov lines \cite{Philipsen}.
Note that an operator dependence might still persist within this
definition \cite{Jahn:2004qr}.

%
\begin{figure}[t]
\center{
  \psfrag{A}[c]{$Q_1$}
  \psfrag{B}[c]{$Q_2$}
  \psfrag{C}[c]{$Q_3$}
  \epsfig{file=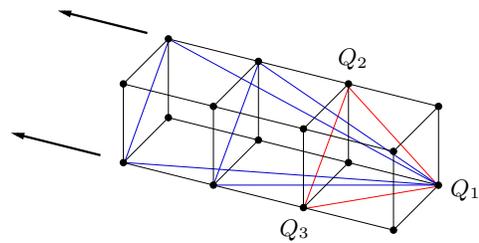, height=3cm}
  \caption{\label{fig_3corr}Calculation of the three point correlation
  function of the Polyakov loop.} 
  
}
\end{figure} 

\subsection{Calculation technique}
\label{calc_technique} 

We first fix the notation for the geometries of the three quark system. 
In fig.~\ref{fig_3eck} we show three quarks $Q_i$ forming a triangle and their
distances $R_{ij}$, where $i,j=1,2,3$. The perimeter of the
triangle is then simply given by
\begin{equation}
  P=\sum_{i<j} R_{ij}.
 \label{eq:perimeter}
\end{equation}
This is also the total length of a
$\Delta$-shaped string, i.~e.~a string connecting the three quarks along the
edges of the triangle. Another possible string shape is a Y-shaped string,
where the flux tube emanetes from each of the three quarks and has a junction at the
Fermat point $F$ of the triangle. The total length of such a Y-shaped string is 
\begin{equation}
  L=\left[ \frac{1}{2}\sum_{i<j}R_{ij}^2 +2\sqrt{3}A_\Delta\right]^{\frac{1}{2}}, 
 \label{eq:ystringlength}
\end{equation}
where $A_\Delta$ is the area of the triangle \cite{Bornyakov:2004uv}. The inner angles at the
vertices of the triangle are assumed to be smaller than $\frac{2\pi}{3}$, which
is the case for all triangles considered in this work. The angle between any two arms
of the Y-shaped flux tube is always $\frac{2\pi}{3}$.
For equilateral triangles we have the simple relation $P=\sqrt{3}L$.
In this work we examine only isoscele triangles, 
where we set $R_{12}=R_{13}=R_s$
to be the equally long edges and $R_{23}=R_b$ to be the length of the base edge.
For equilateral triangles we have $R_{ij}=R$ for all $i,j=1,2,3$.       

The three point correlation functions of the Polyakov loop are now 
obtained as follows (see fig.~\ref{fig_3corr}). 
At the positions of the $Q_i$, we calculate the correlation functions 
\eqref{eq:sing_frenergie} - \eqref{eq:dek_frenergie},
\eqref{eq:average_QQQ} and compute the average of the correlation function 
for those $Q_i$ with the same $\{R_{ij}\}$. 
We obtain a new set of $\{R_{ij}\}$ by 
holding $Q_1$ fixed, whereas the two other vertices of the triangle $Q_2$ and $Q_3$ are moved 
simultanously one point of the lattice in one direction (here: to the left).  
The base edge $R_b$ connecting these to points
preserves thereby its length, which is  $R_b/a=n\sqrt{2}$, where  $a$
denotes the lattice spacing and $1\le n<\frac{N_\sigma}{2}$ is an integer and
describes the number of elementary cells the procedure starts with. 
The two other edges have equal lengths $R_s/a=\sqrt{m^2+n^2}$, where $m$ is another
integer which runs between $n\le m<\frac{N_\sigma}{2}$ for every $n$.
Therefore for every $n$ we obtain one equilateral ($n=m$) and several
isosceles ($n<m< \frac{N_\sigma}{2}$) triangles. 
We start with $n=1$ and repeat the procedure until $n=\frac{N_\sigma}{2}-1$. 
We apply this method in both directions of all three spatial dimensions
before sweeping $Q_1$ over the entire spatial lattice.
We will denote the configuration averages of the correlation functions \eqref{eq:sing_frenergie} - \eqref{eq:dek_frenergie} and \eqref{eq:average_QQQ} by $C^s_{QQQ}$ with $s=1,8,10,\mbox{av}$ from now on.

\begin{figure}[t]
  \epsfig{file=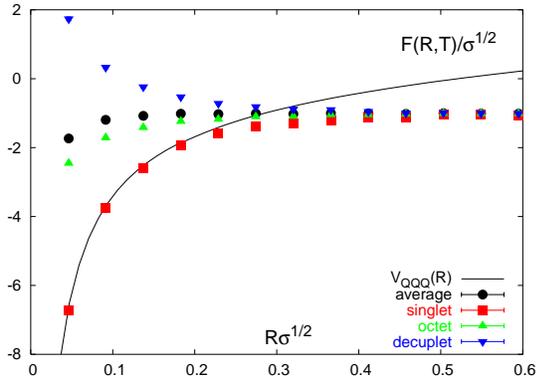,width=7.5cm}
\caption{\label{plc31_ochannels}
     Free energies in different color channels obtained on a $32^3\times 8$ lattice at
     $T/T_c=6$ from simulations in quenched QCD. 
   }  
\end{figure}

\subsection{Renormalization}
\label{renorm}

We construct all correlation functions using renormalised thermal Wilson lines $L^{\text{ren}}(T)$, which are obtained from the bare ones calculated on lattices with temporal extend $N_\tau$ through multiplicative renormalisation,
\begin{equation}
  L^{\text{ren}}(T)=\left( Z(g^2) \right)^{N_\tau} L(g^2,N_\tau),
\end{equation}
where $Z(g^2)$ is the multiplicative renormalisation constant determined in
\cite{Kaczmarek:2002mc}, $g$ is the bare coupling and $N_\tau$ is the temporal extent of the lattice $L$ is
calculated on.
The renormalization constants only depend  on the bare coupling
\cite{Kaczmarek:2005uv,Kaczmarek:2007jw} (and in addition on the bare quark masses in
full QCD) and furthermore are equivalent at zero and finite temperature
\cite{Kaczmarek:2007pb}.
The renormalised three point correlation function of the thermal Wilson lines,
$C^{\text{ren}}(T)$, is then obtained from the bare one, $C_{QQQ}(g^2,N_\tau)$, by
\begin{equation}
  C_{QQQ}^{\text{ren}}(T)=\left( Z(g^2) \right)^{3N_\tau} C_{QQQ}(g^2,N_\tau).
\end{equation}

\section{Results in Pure Gauge}
\label{puregauge}

In this section we will analyze the behavior of three quark free energies in different color channels
above $T_c$. We compare results with
the perturbative expressions obtained in section \ref{perttheory} as well as with results obtained for $QQ$-systems in different
color channels.
Data for the $QQ$ free energies have been taken from \cite{Doring:2007uh}.
Below $T_c$ we examine the string shape of the baryonic system and its structure in the different color channels.

\subsection{Color Channels}
\label{pg_color_channels}

\begin{figure}
  \epsfig{file=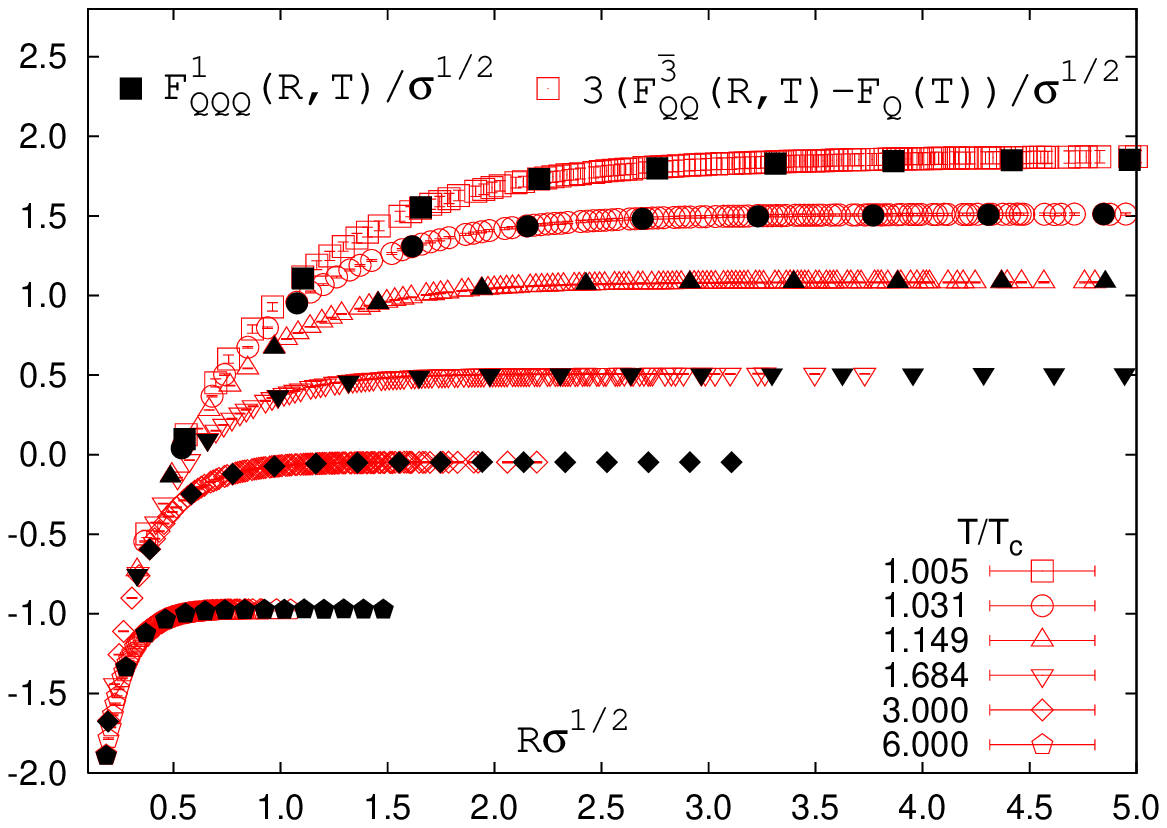,width=7.5cm}
  \caption{\label{plc31}
   $F^1_{QQQ}(R,T)$ and $3\left( F^3_{QQ}(R,T)-F_{Q}(T)\right)$ above $T_c$ versus $R$, the edge length $R$ of the equilateral
triangles and $QQ$-distance, respectively.}  
\end{figure}

In fig.~\ref{plc31_ochannels} we show the free
energies of three quark systems in different color
channels and the average free energy for the $QQQ$-system for equilateral
triangles of edge length $R$ calculated on a $32^3\times 8$ lattice at $T/T_c=6$.   
One can see clearly, that the singlet is strongly, the octet weaker attractive and the
decuplet repulsive in agreement with the perturbative findings presented in sec.~\ref{perttheory}.        
For large $R$ at a given temperature, all free energies in the different color channels 
approach a common value, i.~e.~the three quarks are screened independently of their color orientation.   
The singlet free energy becomes temperature independent at small distances and
coincides with the baryonic $T=0$ potential, $V_{QQQ}(R)$, (see also fig.~\ref{plc31}), which is related to the
quark-antiquark potential at vanishing temperature by the ratio of the different Casimir operators, 
i.~e.~$V_{QQQ}(R)=\frac{3}{2}V_{Q\bar Q}(R)$ for $R\Lambda_{QCD}\ll 1$.
We obtain similiar results for all other temperatures above $T_c$.
We will discuss the screening of octet and decuplet free energies at small distances in sec.~\ref{sec:screening}.
%
%
%
%
\begin{figure}
  \epsfig{file=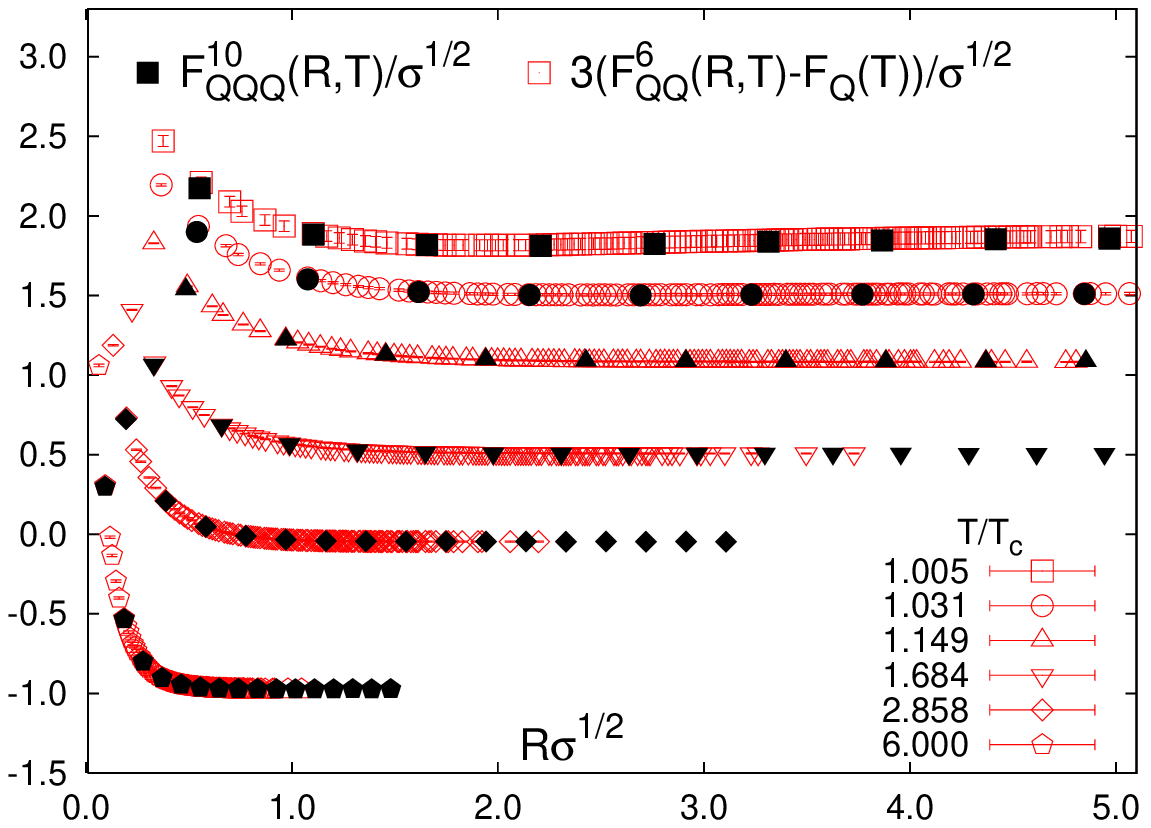,width=7.5cm}
  \caption{\label{plc310}
   $F^{10}_{QQQ}(R,T)$ and $3\left(F^6_{QQ}(R,T)-F_{Q}(T)\right)$ above $T_c$ versus $R$, the edge length $R$ of the equilateral
triangles and $QQ$-distance, respectively.}
\end{figure}
\begin{figure}
  \epsfig{file=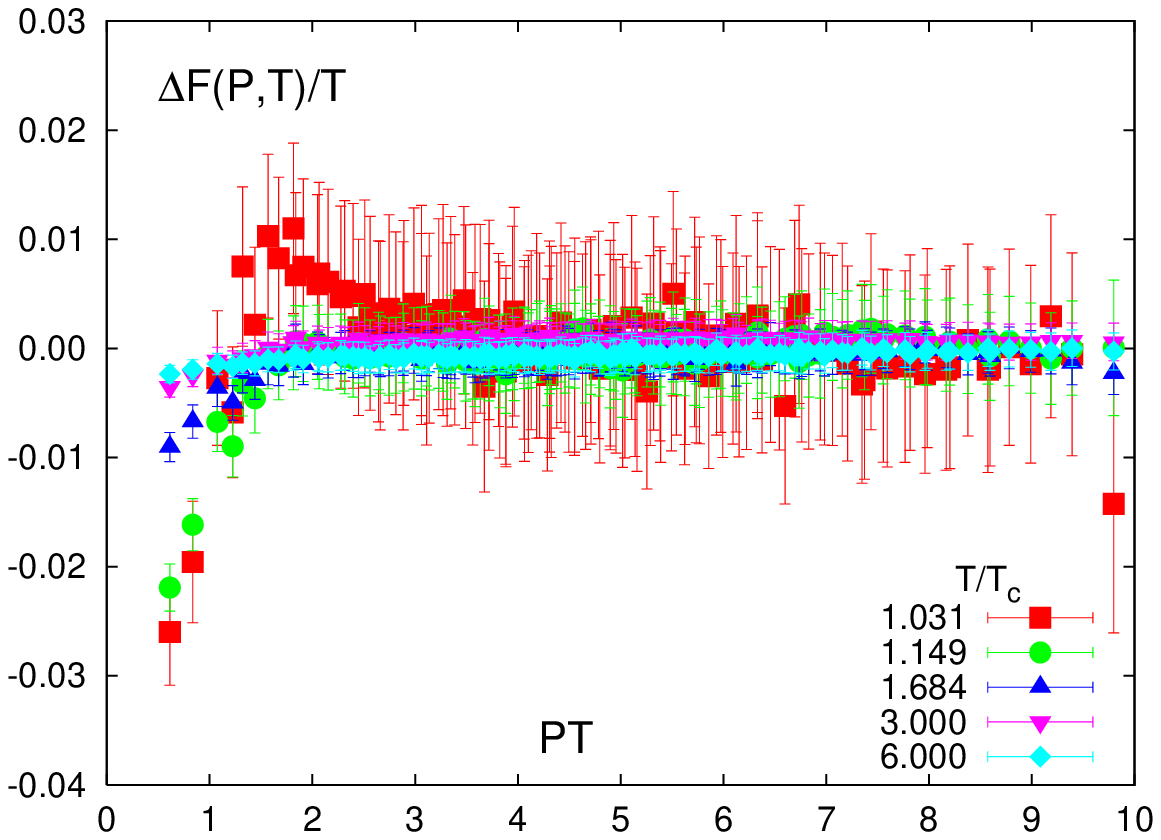,width=7.5cm}
  \caption{\label{plc31_div}
   $\Delta F(P,T)=F^{1}_{QQQ}(P,T)-\sum_i F^{\bar 3}_{QQ}(R_{ij},T) + 3 F_{Q}(T)$ above $T_c$ 
   versus the perimeter $P$ for all geometries calculated.
  }
  
\end{figure}
\begin{figure}
  \epsfig{file=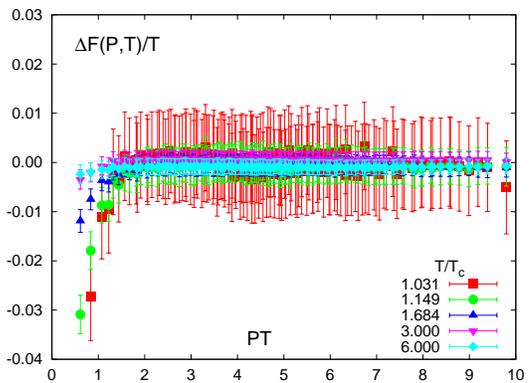,width=7.5cm}
  \caption{\label{plc310_div}
   $\Delta F(P,T)=F^{10}_{QQQ}(P,T)-\sum_{i<j} F^{6}_{QQ}(R_{ij},T) + 3 F_{Q}(T)$ above $T_c$ 
   versus the perimeter $P$ for all geometries calculated.
  }
  
\end{figure}

\subsection{Free energies of equilateral geometries above $T_c$}

We now compare the free energies of the $QQQ$-system with the free energy of the
$QQ$-system above $T_c$.
In fig.~\ref{plc31} we show $F^1_{QQQ}(R,T)$
and $3F^{\bar 3}_{QQ}(R,T)- 3F_Q(T)$ versus the edge length $R$ of the equilateral
triangles and the $QQ$ distance, respectively.
According to (\ref{eq:fre_QQQ_singlet}) the second term is expected to be equal to the $QQQ$-singlet free energy at least 
at small distances, where  genuine
three body forces are negligible. 
For all temperatures above $T_c$ we indeed see that $F^1_{QQQ}(R,T)$ and $3F^{\bar 3}_{QQ}(R,T)
- 3F_{Q}(T)$ 
coincide throughout the entire distance interval.
In fig.~\ref{plc310} we show analogously  $F^{10}_{QQQ}(R,T)$
and $3  F^{6}_{QQ}(R,T)- 3F_{Q}(T)$. Again we observe that both observables
do coincide. 
We obtain similar plots for $F^8_{QQQ}(R,T)$ in accordance with (\ref{eq:fre_eight}).
\begin{figure}
  \epsfig{file=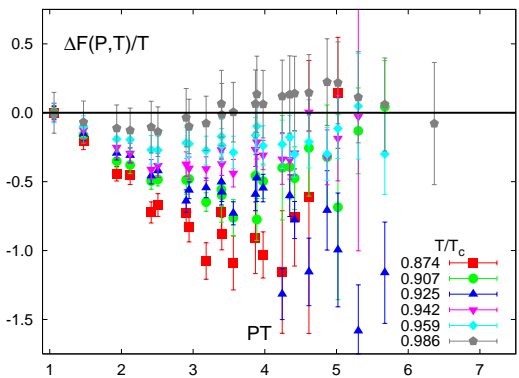,width=7.5cm}
  \caption{\label{plc31_div_u}
   $\Delta F(P,T)=F^{1}_{QQQ}(P,T)-\sum_{i<j} F^{\bar 3}_{QQ}(R_{ij},T)$ below $T_c$ 
   versus the perimeter $P$ for all geometries calculated.
    $\Delta F(P_{\text{min}},T)$ has been set to zero, where $P_{\text{min}}$
   is the smallest perimeter calculated.
  }
\end{figure}

\subsection{Free energies of isosceles geometries above $T_c$}

In order to
test whether the simple relation between free energies of a 3 quark system and the free energies of a 2 quark system plus self energy terms also holds for other geometries and temperatures above $T_c$, we
calculate the difference $\Delta F(P,T)=F^{1}_{QQQ}(P,T)-\sum_{i<j} F^{\bar 3}_{QQ}(R_{ij},T)
+ 3 F_{Q}$. This is shown  in fig.~\ref{plc31_div}. If (\ref{eq:fre_QQQ_singlet}) 
also holds for these geometries, then $\Delta F(P,T)$ 
should vanish, which is fullfilled to a very good degree for all
perimeters except the smallest ones. 
Again the same holds true also for the  $QQQ$-decuplet free energy as is evident from
fig.~\ref{plc310_div}.
This is possibly due to the effect, that the self energy of the system is no longer the sum of the self energy of the individual static quarks at short distances but rather that of a single static quark in the corresponding color representation (see sec.~\ref{sec:screening}).

Having now established that (\ref{eq:fre_QQQ_singlet}) and
(\ref{eq:fre_QQQ_decuplet}) hold for all isoscele geometries above $T_c$
calculated in this work, 
it is clear 
that the screening length of the three quark
system in the singlet state is the same 
as that of the antitriplet diquark state, which itself has been shown to be
equal to that of the $Q\bar Q$ singlet state \cite{Doring:2007uh}, therefore reflecting
the properties of the thermal medium rather than that of a particular hadronic system.

Furthermore, we can apply the relations used in \cite{Doring:2007uh} to calculate entropy and internal energy from free energies of a $QQ$ system also for the $QQQ$ system.
The entropy in the color singlet channel is defined by
\begin{equation}
  \label{QQQ_entropy}
  S^{1}_{QQQ}(P,T)=\frac{\partial F^{1}_{QQQ}(P,T)}{\partial T}
\end{equation}
and the internal energy in this channel is 
\begin{equation}
  U^{1}_{QQQ}(P,T)=-T^2\frac{\partial F^{1}_{QQQ}(P,T)/T}{\partial T}.
\end{equation}
We then obtain for the entropy
\begin{equation}
  \lim_{P\to 0}S^{1}_{QQQ}(P,T) = 0,
\end{equation}
i.~e.~the entropy contribution in the free energy of the $QQQ$ singlet color
channel vanishes at small distances. 
Moreover, we find for the
internal energy, that
\begin{equation}
  \lim_{P\to 0}U^{1}_{QQQ}(P,T) = \lim_{P\to 0} V_{QQQ}(P),
\end{equation}
where $V_{QQQ}(P)=\frac{1}{2}\sum_{i<j}V_{Q\bar Q}(R_{ij})$ and $V_{Q\bar Q}(r)$ is the quark-anti-quark potential at $T=0$.
This means that the internal energy becomes $T$-independent for small perimeters and thus,
together with (\ref{QQQ_entropy}), the singlet free energy itself is
$T$-independent for small perimeters, as was already seen in sec.~\ref{pg_color_channels}.
This behavior is already known for $Q\bar
Q$ singlet free energies and reflects the fact that the baryonic system is less
and less effected by the surrounding thermal medium when going to smaller and
smaller perimeters.

\subsection{Free energies below $T_c$}

\begin{figure}[t]
  \epsfig{file=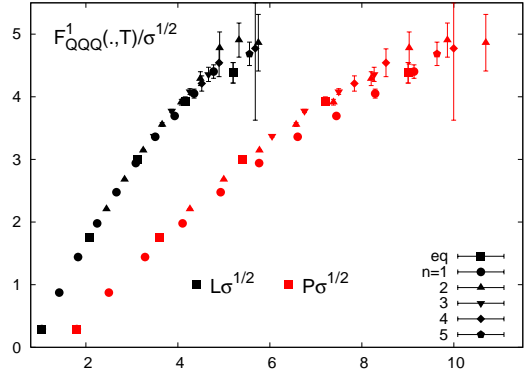,width=7.5cm}
  \caption{\label{plc31_ystring}
   $F^{1}_{QQQ}(.,T)$ at $T/T_c=0.925$ 
   versus the Y-string length $L$ and the perimeter $P$ for
   equilateral geometries (eq) and geometries with base length  $R_b\sqrt{\sigma}=n \sqrt{\sigma}\sqrt{2}$, where $n$ is
   given in the legend.
  }
\end{figure}
We now examine the $QQQ$ free energies below $T_c$. 
We start by looking at the relation
between  the $QQQ$-singlet and the $QQ$-antitriplet free energies.
If $F^{1}_{QQQ}$ can be expressed in terms of 
the sum of $F^{\bar 3}_{QQ}(R_{ij},T)$ also below $T_c$, i.~e.~if a $\Delta$-ansatz
for the flux tube shape together with the same string tension holds, 
then $F^{1}_{QQQ}$ is a function of the perimeter only  and 
$\Delta F(P,T)=F^{1}_{QQQ}(P,T)-\sum_{i<j}
F^{\bar 3}_{QQ}(R_{ij},T)$ should be equal to a $T$-dependent constant $k(T)$ for
all $P$.
In fig.~\ref{plc31_div_u} we show $\Delta F(P,T)/T$ for $T<T_c$ versus the perimeter, $P$, for different geometries. Here we have set $\Delta
F(P_{\text{min}},T)=0$, where $P_{\text{min}}$ is the smallest perimeter
calculated. 
We can see clear deviations of $\Delta F(P,T)$ to smaller values, most strongly
for the lowest temperatures, deviations from zero becoming smaller with growing
temperature. For $T/T_c=0.986$ we have $\Delta F(P,T)\approx 0$. 
Hence, $F^{1}_{QQQ}(P,T)$ can not  be expressed in terms of 
the sum of $F^{\bar 3}_{QQ}(R_{ij},T)$ except close to the critical
temperature. 
We are left with two possibilities now.
First, we could still have a $\Delta$-shaped flux tube, but with a different
string tension than that observed for the $QQ$-antitriplet. Since we see
$\Delta F(P,T)<0$, we would expect the string tension to be smaller than in the $QQ$-antitriplet.
In this case the
perimeter $P$ would still be the right distance measure for the
$QQQ$-singlet free energy, i.~e.~$F^{1}_{QQQ}$ should be a smooth function of
$P$ for all geometries.
Or, secondly, the flux tube is Y-shaped and $L$ is the right  distance
measure. In this case, $F^{1}_{QQQ}$ should be a smooth function of
$L$ for all geometries in which a flux tube can form.    
\begin{figure}[t]
  \epsfig{file=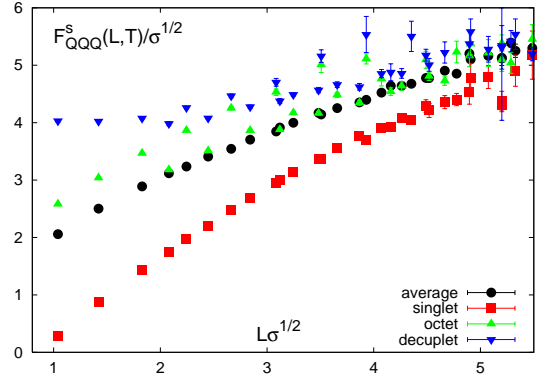,width=7.5cm}
  \caption{\label{plc3_cc_u}
    Free energies of the color channels 
   versus the Y-string length $L$ at $T/T_c=0.925$  for all geometries calculated.
  }
  
\end{figure}

To elaborate more on the shape of the flux tube, 
we take a closer look at the
$QQQ$-singlet free energy at a particular temperature below $T_c$. We analyze the free energy for different
geometries as a function of $L$ and $P$. 
For equilateral triangles a
simple geometrical relation exists between the length of a Y-shaped flux tube $L$ and the length
of a $\Delta$-shaped flux tube $P$, which is $P=\sqrt{3}L$.
Hence for equilateral geometries the $QQQ$-singlet free energy is a smooth function in
both Ans\"atze. 
For more general geometries like the isoscele triangles
we calculated, no simple relation between $L$ and $P$ exists. 
This may help to clarify the situation more directly. 
Therefore we plot the $QQQ$-singlet free energy at $T/T_c=0.925$ versus
the Y-string length $L$ and the perimeter $P$. This is shown in fig.~\ref{plc31_ystring}
for equilateral geometries and also for isoscele geometries up to $n=5$ (see sec.~\ref{calc_technique}). 
We observe that the $QQQ$-singlet free energy is
indeed a smooth function of $L$ for all geometries calculated.
On the other hand, when the free energy is plotted versus 
the perimeter $P$ different branches become visible depending on the
geometry. This is most prominent for the $n=1$ triangles.     
Therefore we find strong hints, that the shape of the flux tube in the $QQQ$-singlet
system is indeed that of the Y-ansatz and its string length $L$ is the right
distance measure for the system.  

Having established this, we take a look at the other color channels below
$T_c$ and analyze whether they display a smooth behavior as function of $L$ as well.
In fig.~\ref{plc3_cc_u} we show the different color channels of the $QQQ$-free
energy at $T/T_c=0.925$ over the Y-string length $L$. 
We see again that the singlet is the most attractive channel followed by the
average free energy, which is also a smooth function of $L$.
The octet channel is still attractive, but weaker so than the average free energy.
The decuplet free energy is attractive for large $L$ but becomes flat at
smaller $L$, possibly hinting at a turnover and at a repulsive behavior at even smaller
$L$. 
Both the decuplet and the octet channel are not smooth functions over either
distances $L$ and $P$, but become volatile for $L\sqrt{\sigma}\gsim 2$.
This suggests that no flux tube forms in these two channels, besides that the
octet free energy shows an overall attractive behavior.

\begin{figure}[t]
  \epsfig{file=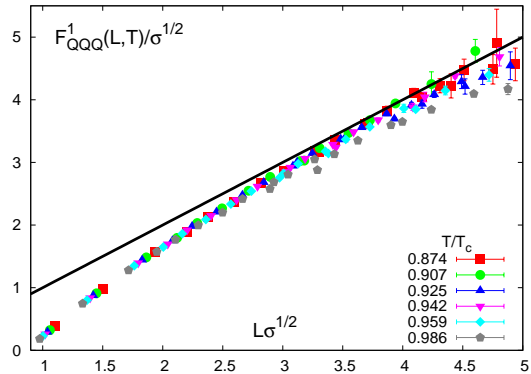,width=7.5cm}
  \caption{\label{plc31_ren_u}
   $F^{1}_{QQQ}(L,T)$ below $T_c$ 
   versus the Y-string length $L$ for all geometries. 
  }
\end{figure}

Finally, we examine more closely the temperature dependence of $F^{1}_{QQQ}$.
In fig.~\ref{plc31_ren_u} we show the renormalised $QQQ$-singlet free energy versus $L$ for
all temperatures below $T_c$.
We observe $F^{1}_{QQQ}(L,T)$ to nearly
coincide at distances $L\sqrt{\sigma}\lsim 4$ for all temperatures $T/T_c\le
0.959$. We see that at large distances the free energies approach $F(L,T)=\sigma L$ represented by the black line in fig.~\ref{plc31_ren_u}, where $\sigma$ denotes the string tension in a $Q\bar Q$-system. 
The phenomenon of an almost $T$-independent string tension has also been
observed in \cite{Bornyakov:2004yg}, using a fit ansatz in full QCD. 
As was already seen in fig.~\ref{plc31_div_u}, at $T/T_c=0.986$ we observe deviations to smaller values.
This might be due to flux tube broadening setting in near $T_c$, leading to an overlapping of the three branches of the Y-shaped flux tube. Filling the space between the three static quarks the separation dependence of the singlet free energy is effectively described by the $\Delta$-Ansatz at these temperatures.

\section{Results in 2-flavor QCD}
\label{fullqcd}

In this section we analyse the $QQQ$-free energies obtained on a $16^3\times
4$ lattice in 2-flavor QCD with a bare quark mass of $\frac{m}{T}=0.4$ . For further details, see sec.~\ref{details}. 
Data for the $QQ$ free energies is again taken from \cite{Doring:2007uh}.

\subsection{Color Channels}
\begin{figure}[t]
  \epsfig{file=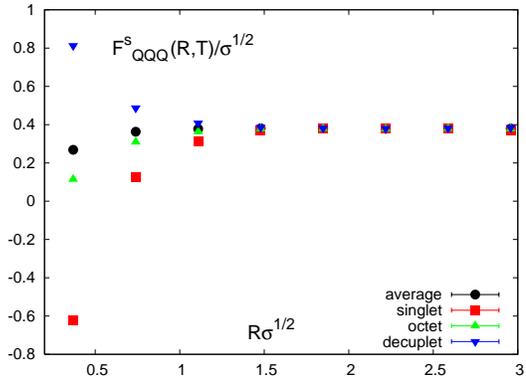,width=7.5cm}
  \caption{
    Free energies in different color channels in 2-flavor QCD from calculations on a $16^3\times 4$ lattice at
    $T/T_c=1.99$.
  }
  \label{plc31_1.99_fqcd}
\end{figure}

In fig.~\ref{plc31_1.99_fqcd} we show the free energies of three quark systems with equilateral
geometries in different color channels at $T/T_c=1.99$.  
Like in the pure gauge case, we observe the singlet to be strongly, the octet
weaker attractive and the decuplet to be repulsive. For large distances $R$ the
free energies in all color channels approach a common
value. Again, we obtain similar results for all $T>T_c$.

\subsection{Singlet free energy}

\begin{figure}[t]
  \epsfig{file=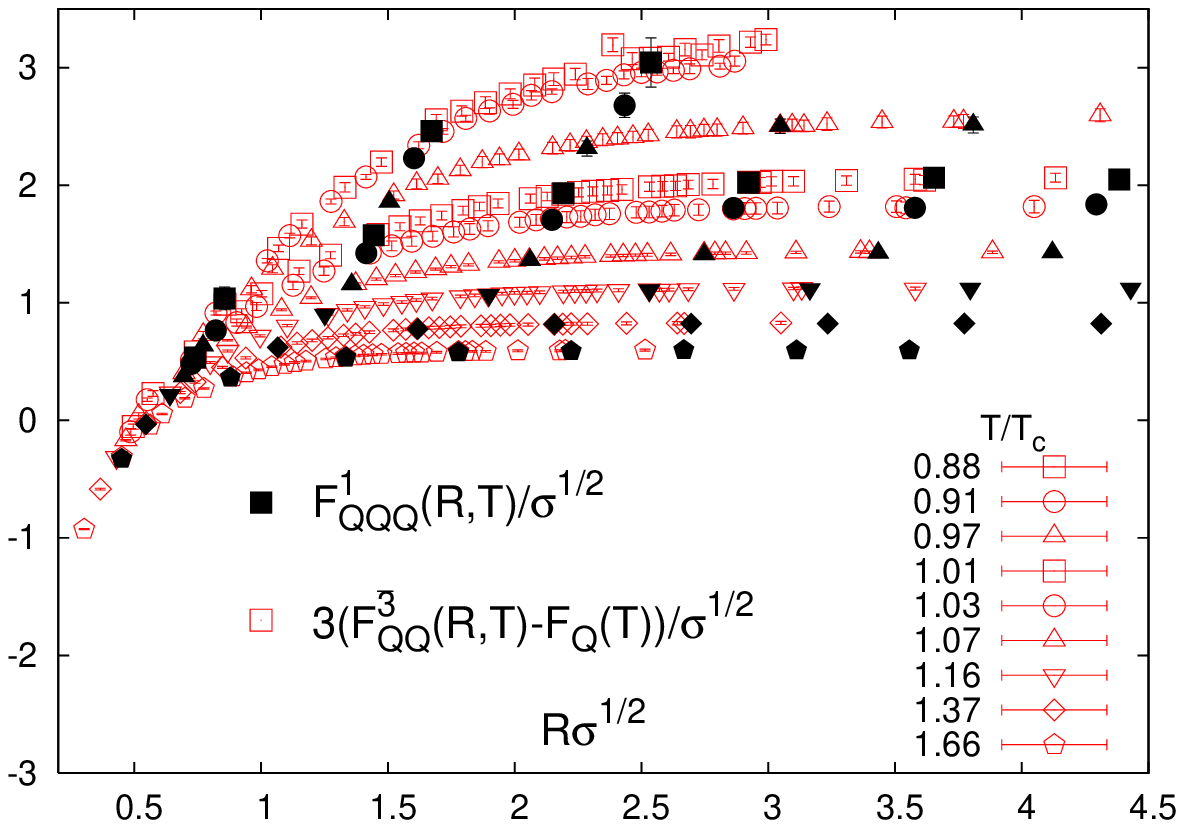,width=7.5cm}
  \caption{
   $F^1_{QQQ}(R,T)$ and $3(F^{\bar 3}_{QQ}(R,T)-F_{Q}(T))$ versus $R$, the edge length of the equilateral
triangles and $QQ$-distance, respectively.
  }
  \label{plc31_fqcd}
\end{figure}

We now compare the free energy of the $QQQ$-singlet with the free energy of the
$QQ$-antitriplet for 2-flavor QCD. 
In fig.~\ref{plc31_fqcd} we show $F^1_{QQQ}(R,T)$
and $3F^{\overline{3}}_{QQ}(R,T)-3F_{Q}(T)$ versus the edge length $R$ of the equilateral
triangles and $QQ$ distance, respectively. 
As in pure gauge theory, both curves coincide for all
temperatures above $T_c$. Thus we find the $QQQ$-singlet free energy of
equilateral geometries can be
described as the sum of three $QQ$-antitriplet free energies plus self energy
terms as well in 2-flavor QCD above $T_c$.

\begin{figure}
  \epsfig{file=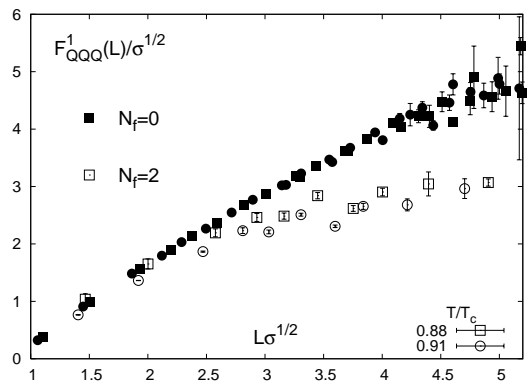,width=7.5cm}
  \caption{\label{plc31_strbrk}
    Comparison of $F^{1}_{QQQ}(L,T)$ in 2-flavor QCD (open symbols) and SU(3) pure gauge
  theory (filled symbols) for two temperatures below $T_c$. 
  }
  
\end{figure}

Below $T_c$ we expect to see string breaking also in the $QQQ$-singlet free
energy, the breaking mechanism being more involved than in the $Q\bar Q$ case \cite{Bornyakov:2004yg}.
Indeed, except at $T/T_c=0.97$, where both quantities agree, we see
deviations for the $QQQ$-singlet from the simple relation to the $QQ$ free energies.
In fig.~\ref{plc31_strbrk} we compare $F^{1}_{QQQ}(L,T)$ in 2-flavor QCD
 ($T/T_c=0.88,0.91$) and pure gauge
theory ($T/T_c=0.874,0.907$) for all geometries calculated. 
Here the pure gauge data have been obtained from simulations on a
$32^3\times 4$ lattice. 
The $N_f=2$ free energies start to deviate from the pure gauge result already at short distances to
smaller values and eventually become flat. Specifying a definite value for the
string breaking distance is quite difficult given the present
data. Nevertheless we can give a rough estimate for the distance at which the
pure gauge free energies assumes the
asymptotic value of the 2-flavor free energies, which is $L\sqrt{\sigma}\approx
3$. For higher $T$, this distance becomes smaller. 
The string breaking distance for the $Q\bar Q$ singlet free energies has been
determined in \cite{Kaczmarek:2005uv} as the distance where the $T=0$ potential
assumes the large distance asymptotic value of the free energy.
This lead to values between $2.8/\sqrt{\sigma}$ 
for $T/T_c=0.874$ and $1.7/\sqrt{\sigma}$ close to $T_c$.


\subsection{Decuplet free energies}

\begin{figure}
  \epsfig{file=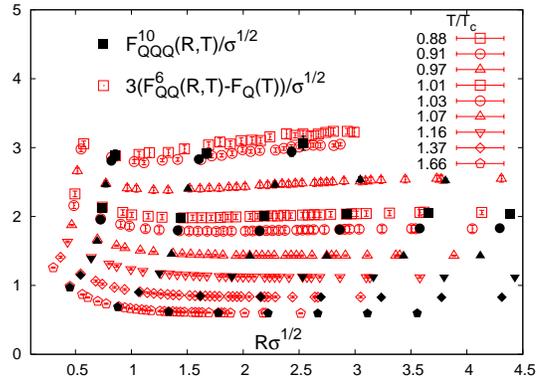,width=7.5cm}
  \caption{
   $F^{10}_{QQQ}(R,T)$ and $3(F^6_{QQ}(R,T)-F_{Q}(T))$ versus $R$, the edge length of the equilateral
triangles and $QQ$-distance, respectively.
  }
  \label{plc310_fqcd}
\end{figure}

For the $QQQ$-decuplet free energies we show $F^{10}_{QQQ}(R,T)$ and 
$3(F^6_{QQ}(R,T)-F_{Q}(T))$ for equilateral geometries in
fig.~\ref{plc310_fqcd}.
Like in pure gauge theory, 
we observe both quantities  to coincide for
all temperatures above $T_c$ and distances $R$. 
The situation below $T_c$ is analogous to the singlet case. At the temperature
closest to $T_c$ both quantities coincide for all $R$, at  smaller $T$ we observe
deviations to smaller values for $F^{10}_{QQQ}(R,T)$. 

Hence we see, that the singlet (decuplet) $QQQ$ free energy can be
described as a sum of antitriplet (sextet) $QQ$ free energies and the self
energy contributions of the three quarks above $T_c$    
(Eqs (\ref{eq:fre_QQQ_singlet}) and
(\ref{eq:fre_QQQ_decuplet})) also in 2-flavor QCD.
Again, we obtain similar plots for $F^8_{QQQ}(R,T)$ in accordance with (\ref{eq:fre_eight}).

%
%
%

\section{Screening of octet and decuplet free energies}
\label{sec:screening}

As was shown in \cite{Doring:2007uh}, the  screening of diquark free energies 
in the anti-triplet and sextet representation 
at small distances becomes identical to the  
screening of a single fermion in a color anti-triplet and sextet representation, respectively.
This holds
for all temperatures in 2-flavor QCD and in the deconfinement phase of SU(3) pure 
gauge theory. 

We will now show, that analogous statements are true in the deconfinement phase 
for octet free energies of the $Q\overline{Q}$-system  
and for octet and decuplet free energies of the $QQQ$-system, as we have proposed in sec.~\ref{perttheory}.

\begin{figure}[t]
  \epsfig{file=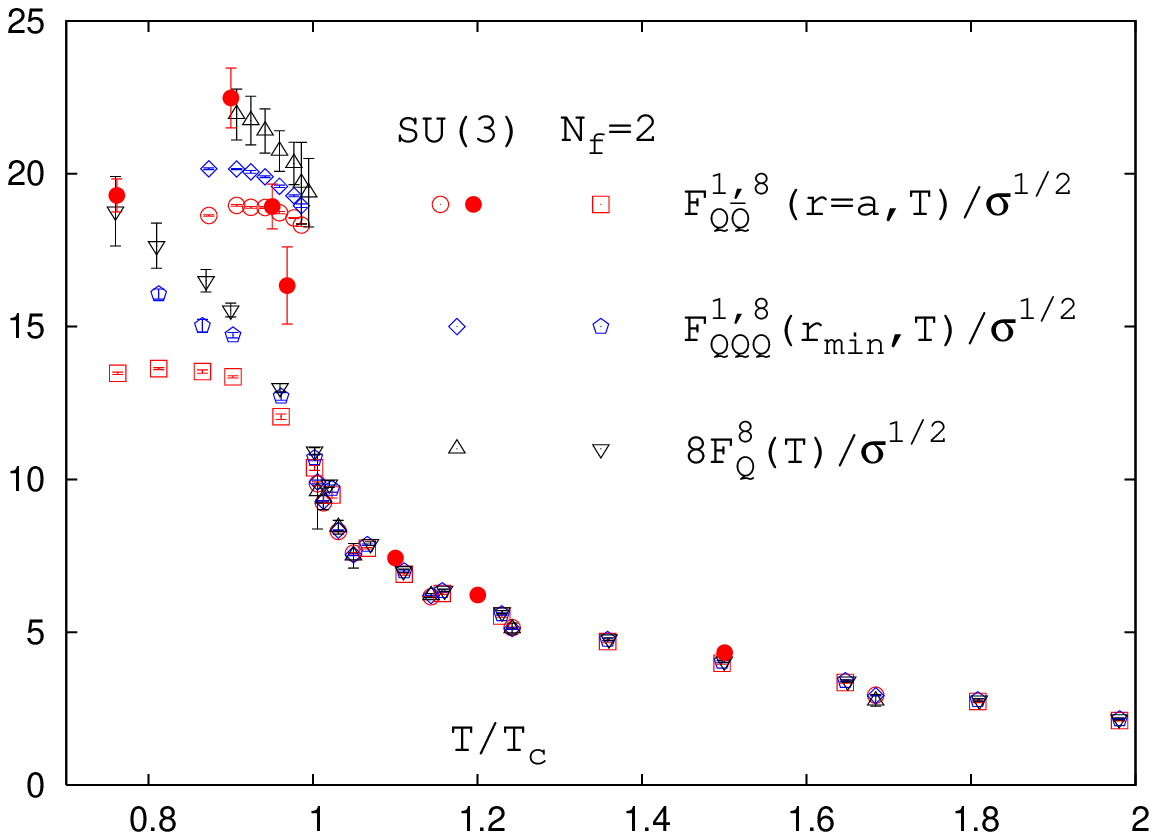,width=7.5cm}
  \caption{\label{plc38_screen}Comparison of singlet and octet free energies at small distances 
with the free energy of a static fermion in the adjoint representation. 
We have used $F^{1,8}_{Q\overline{Q}}(R=a,T)=F^{1}_{Q\overline{Q}}(R=a,T)+8F^{8}_{Q\overline{Q}}(R=a,T)$ and 
$F^{1,8}_{QQQ}(R_{\text{min}},T)=-2F^{1}_{QQQ}(R_{\text{min}},T)+8F^{8}_{QQQ}(R_{\text{min}},T)$.
This particular combinations of $Q\overline{Q}$- and $QQQ$-free energies shown eliminate 
the short distance Coulomb terms contributing to them according to (\ref{F8}) - (\ref{F8b}). 
Open symbols: $N_\tau=4$ data, closed symbols: $N_\tau=8$ data. 
For the $QQQ$-free energies we used equilateral geometries with edge length $R_{\text{min}}=\sqrt{2}a$.
  }
\end{figure}

When combining singlet and octet free energies or singlet and decuplet free energies in
proportion to  the corresponding Casimir factors (cf.~Tab.~\ref{prefactor_tab}) such that
the repulsive and attractive Coulombic contributions cancel, we expect 
to find 
\begin{eqnarray}
\label{F8}
&&\lim_{R\rightarrow 0} \left(F^{(1)}_{Q\overline{Q}}(R,T) + 8F^{(8)}_{Q\overline{Q}}(R,T)\right)\\\non &&\\
& = &\label{F8a} \lim_{r\rightarrow 0} \left(-2F^{(1)}_{QQQ}(R,T) + 8F^{(8)}_{QQQ}(R,T)\right)\\\non &&\\
\label{F8b}
& = &  8 F^{(8)}_Q (T)\\\non
\end{eqnarray} 
and 
\begin{equation}
\lim_{R\rightarrow 0} \left(F^{(1)}_{QQQ}(R,T) + 
2F^{(10)}_{QQQ}(R,T)\right)=2 F^{(10)}_Q (T)
\label{F10}
\end{equation} 
where $R$ stands for the edge length of the equilateral geometries in the 
case of $QQQ$ free energies; 
 $F^{(8)}_Q (T)$ and $F^{(10)}_Q (T)$ are the free energies of a static quark in a color 
octet and decuplet representation, respectively. The latter are given by the corresponding Polyakov loop
expectation values,
\begin{equation}
{\rm e}^{-F^{(D)}_Q /T} = \left\langle L_D(\bx)\right\rangle,
\label{Fs}
\end{equation} 
where $D=8,10$ and
\begin{eqnarray}
\non\\
L_8(\bx) & = & \frac{1}{8}\left(9\vert L(\bx)\vert^2-1\right)\\&&\non\\
L_{10}(\bx) & = & \frac{1}{10}(27L(\bx)^3- 18\vert L(\bx)\vert^2+1)\\\non
\label{F8,10_aux}
\end{eqnarray} 
are obtained through group theoretical relations. 
The Polyakov loops $\left\langle L_8\right\rangle$ and $\left\langle L_{10}\right\rangle$ 
have been computed in \cite{huebner}. We note here, that these quantities 
-- unlike the Polyakov loop in the fundamental representation -- do not have to vanish 
in the confinement phase of SU(3) pure gauge theory due to their vanishing triality.

\begin{figure}[t]
  \epsfig{file=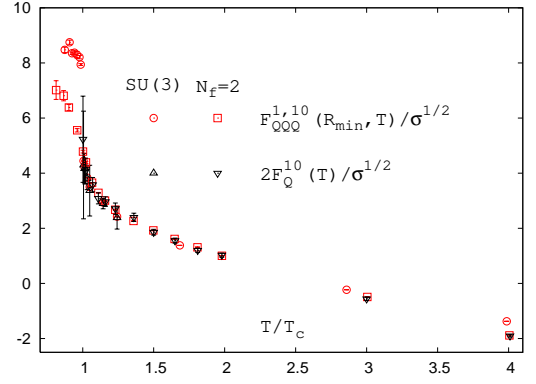,width=7.5cm}
  \caption{\label{plc310_screen}Comparison of singlet and decuplet free energies at small distances with the free energy of a static fermion in the decuplet representation. We have used $F^{1,10}_{QQQ}(R_{\text{min}},T)=F^{1}_{QQQ}(R_{\text{min}},T)+2F^{10}_{QQQ}(R_{\text{min}},T)$. This particular combination of $QQQ$-free energies shown eliminates the short distance Coulomb terms contributing to them accroding to (\ref{F10}). For the $QQQ$-free energies we used $N_\tau=4$ data and again equilateral geometries with edge length $R_{\text{min}}=\sqrt{2}a$.
   }
\end{figure}

(\ref{F8}) - (\ref{F10}) are indeed fulfilled for all temperatures above $T_c$ as can be seen
from Figs.~\ref{plc38_screen} and \ref{plc310_screen}. 
In particular, the combinations of singlet and octet free energies of the 
$Q\overline{Q}$- and the $QQQ$-system  coincide 
in the deconfinement phase (\ref{F8a}), thus showing the screening at small distances 
to depend only on the color representation and not on the particular static quark system under consideration. 
Below $T_c$ deviations become apparent in Fig.~\ref{plc38_screen}. 
The reason probably is, that
the minimal distances used in the present analysis ($R=a$ for the $Q\overline{Q}$-system and $R_{min}=\sqrt{2}a$ 
for the $QQQ$-system respectively)
are still too large at small temperatures to be a good approximation for the 
small distance limit in (\ref{F8}) - (\ref{F10}). In fact, the $N_\tau=8$ data, where the lattice spacing $a$ is only half as large 
at the same value of the temperature, agree much better at least close to $T_c$.

\section{Conclusions}
\label{conclusion}

We have calculated the free energy of static three quark 
systems in different color channels 
in the quenched approximation and in 2-flavor QCD at finite temperature. 
We have shown that above the critical temperature the singlet and decuplet 
free energies of the three quark system can be described by the sum of the free energies of the
corresponding diquark system plus self energy contributions.
Therefore the screening of the singlet $QQQ$ free energies
is comparable to that of the $QQ$ free energies reflecting the screening
properties of the thermal medium.  
Below $T_c$ we found evidence for a Y-shaped
flux tube in the quenched approximation for the singlet and color average
channel. The string tension agrees with that deduced from $Q\bar Q$-systems at $T=0$ and has been found to be 
independent of temperature for $T/T_c\lsim 0.959$ in the singlet color
channel. 
In 2-flavor QCD we observed string breaking in the $QQQ$ singlet free energies 
at distances around $3/\sqrt{\sigma}$.
Moreover, we were able to show that at short distances the screening properties of octet free energies of the $Q\bar Q$- and the $QQQ$-system and the decuplet free energies of the $QQQ$-system coincide with those of a single static quark in the corresponding color state above $T_c$.

\section*{Acknowledgment}
This manuscript has been authored under Contract No. DE-AC02-98CH10886 with the U.~S.~Department of Energy.
O.V. and at an early stage K.H. have been supported by the Deutsche
Forschungsgemeinschaft (DFG) under grant GRK 881.


\begin{thebibliography}{99}

\bibitem{Sommer:1985da}
  R.~Sommer and J.~Wosiek,
  Nucl.\ Phys.\  B {\bf 267}, 531 (1986)

\bibitem{Bali:2000gf}
  G.~S.~Bali,
  Phys.\ Rept.\  {\bf 343}, 1 (2001)

\bibitem{Alexandrou:2001ip}
  C.~Alexandrou, P.~De Forcrand and A.~Tsapalis,
  Phys.\ Rev.\  D {\bf 65}, 054503 (2002)

\bibitem{Takahashi:2002bw}
  T.~T.~Takahashi, H.~Suganuma, Y.~Nemoto and H.~Matsufuru,
  Phys.\ Rev.\ D {\bf 65}, 114509 (2002) 


\bibitem{Takahashi:2004rw}
  T.~T.~Takahashi and H.~Suganuma,
  Phys.\ Rev.\  D {\bf 70}, 074506 (2004)

\bibitem{Bornyakov:2004uv}
V.~G.~Bornyakov {\it et al.}  [DIK Collaboration],
Phys.\ Rev.\ D {\bf 70} (2004) 054506



\bibitem{Bornyakov:2004yg}
  V.~G.~Bornyakov {\it et al.},
  Prog.\ Theor.\ Phys.\  {\bf 112}, 307 (2004)


\bibitem{deForcrand:2005vv}
  Ph.~de Forcrand and O.~Jahn,
  Nucl.\ Phys.\  A {\bf 755}, 475 (2005)






\bibitem{Kuzmenko:2000bq}
 D.~S.~Kuzmenko and Y.~A.~Simonov,
   Phys.\ Lett.\ B {\bf 494}, 81 (2000)




\bibitem{Liao:2005hj}
  J.~Liao and E.~V.~Shuryak,
  Nucl.\ Phys.\  A {\bf 775}, 224 (2006)


\bibitem{Nadkarni:1986as}
  S.~Nadkarni,
  Phys.\ Rev.\ D {\bf 34}, 3904 (1986)

\bibitem{Nadkarni:1986cz}
  S.~Nadkarni,
  Phys.\ Rev.\ D {\bf 33}, 3738 (1986)

\bibitem{Cornwall:1996xr}
  J.~M.~Cornwall,
  Phys.\ Rev.\ D {\bf 54}, 6527 (1996)

\bibitem{Doring:2007uh}
  M.~D\"oring, K.~H\"ubner, O.~Kaczmarek and F.~Karsch,
  Phys.\ Rev.\  D {\bf 75}, 054504 (2007)

\bibitem{Kaczmarek:2002mc}
O.~Kaczmarek, F.~Karsch, P.~Petreczky and F.~Zantow,
Phys.\ Lett.\ B {\bf 543}, 41 (2002)

\bibitem{weisz1}
  P.\ Weisz, Nucl.\ Phys.\ B {\bf 212}, 1 (1983)

\bibitem{weisz2}
  P.\ Weisz and R.\ Wohlert, Nucl.\ Phys.\ B {\bf 236}, 397 (1984)

\bibitem{Karsch:2000kv}
F.~Karsch, E.~Laermann and A.~Peikert,
Nucl.\ Phys.\ B {\bf 605}, 579 (2001)



\bibitem{Allton:2002zi}
  C.~R.~Allton {\it et al.},
  Phys.\ Rev.\ D {\bf 66}, 074507 (2002) 

\bibitem{Allton:2003vx}
C.~R.~Allton {\it et al.},
Phys.\ Rev.\ D {\bf 68}, 014507 (2003) 


\bibitem{beinlich}
  B.\ Beinlich, F.\ Karsch and A.\ Peikert, 
  Phys.\ Lett.\ B {\bf 390}, 41 (1997)


\bibitem{Philipsen}
O.~Philipsen,
Phys.\ Lett.\ B {\bf 535}, 138 (2002)

\bibitem{Jahn:2004qr}
  O.~Jahn and O.~Philipsen,
  Phys.\ Rev.\  D {\bf 70}, 074504 (2004)


\bibitem{Kaczmarek:2005uv}
  O.~Kaczmarek and F.~Zantow,
  Eur.\ Phys.\ J.\ C {\bf 43}, 63 (2005)

\bibitem{Kaczmarek:2007jw}
  O.~Kaczmarek, S.~Gupta and K.~H\"ubner,
  PoS(LATTICE 2007)195
  arXiv:0710.2277 [hep-lat].

\bibitem{Kaczmarek:2007pb}
  O.~Kaczmarek,
  PoS C {\bf POD07}, 043 (2007)

\bibitem{huebner}
S.~Gupta, K.~H\"ubner and O.~Kaczmarek,
in preparation







\end{thebibliography}
\end{document}